# Onset of multifragmentation and vaporization in Au-Au collisions.

Supriya Goyal and Rajeev K. Puri*
*Department of Physics, Panjab University, Chandigarh-160014, INDIA*
* email: rkpuri@pu.ac.in

## Introduction

Various phenomena such as incomplete (and complete) fusion-fission at low incident energies as well as multifragmentation or a complete disassembly of the nuclear matter are the typical characteristics seen in a heavy-ion reaction. Lot of experimental and theoretical studies are reported in the literature during last two decades on the breaking of colliding nuclei into pieces of different sizes. A rise and fall in the multiplicity of intermediate mass fragments (IMFs), originating from the projectile was reported by Ogilvie *et al.* [1]. Same trend of rise and fall was also reproduced at incident energies of 250, 400, and 1000 MeV/nucleon [2,3]. This rise and fall was reported to halt at incident energy of 100 MeV/nucleon, where monotonic decrease was reported with impact parameter [3]. At theoretical front, lots of isolated studies are available in the literature that explains the experimental results [4]. No study, however, is available under one umbrella, that gives a view of the reaction dynamics and associated phenomena collectively. Therefore, in the present study, we present a systematic study of the multifragmentation for the reaction of $^{197}$Au+$^{197}$Au over entire incident energy and impact parameter range. Earlier, $^{40}$Ca+$^{40}$Ca reaction was studied [5], but no complete systematic studies are available in the literature for the reaction of $^{197}$Au+$^{197}$Au. The quantum molecular dynamics (QMD) model is used for the present study [6].

## Model

The QMD model simulates the heavy-ion reactions on event by event basis. This is based on a molecular dynamic picture where nucleons interact via two and three-body interactions. The nucleons propagate according to the classical equations of motion:

$$\frac{d\mathbf{r}_i}{dt} = \frac{dH}{d\mathbf{p}_i} \text{ and } \frac{d\mathbf{p}_i}{dt} = -\frac{dH}{d\mathbf{r}_i} \quad , \quad (1)$$

where H stands for the Hamiltonian which is given by

$$H = \sum_i \frac{\mathbf{p}_i^2}{2m_i} + V^{tot} . \quad (2)$$

Our total interaction potential $V^{tot}$ reads as

$$V^{tot} = V^{Loc} + V^{Yuk} + V^{Coul}, \quad (3)$$

where $V^{Loc}$, $V^{Yuk}$, and $V^{Coul}$, respectively, the local (two and three-body) Skyrme, Yukawa, and Coulomb potentials. For details, the reader is referred to Ref. [6].

## Results and discussion

For the present study, we simulated the reactions of $^{197}$Au+$^{197}$Au between 20 and 1000 MeV/nucleon over entire colliding geometry (b/b$_{max}$ = 0-0.98, where b$_{max}$ = $R_1$ + $R_2$; $R_i$ is the radius of projectile or target) using soft equation of state along with energy dependent nucleon-nucleon cross-section. The calculations are performed using minimum spanning tree algorithm with microscopic binding energy cut (MSTB(2.1)) [7]. The freeze-out time is taken to be between 300-500 fm/c. For the low incident energies, such as 20 MeV/nucleon it is 500 fm/c, while for higher incident energies it is 300 fm/c. In Fig. 1, we display the final stage multiplicities of heaviest fragment (<A$^{max}$>), free nucleons, and IMFs (5≤A≤65) as a function of scaled impact parameter (b/b$_{max}$). The size of the heaviest fragment will give us the possibility to look for partial fusion (if any) in the reaction of $^{197}$Au+$^{197}$Au, whereas the emission of the free nucleons will show the disassembly (and hence vaporization) of the colliding matter. From the figure, we clearly see that no partial fusion is seen in $^{197}$Au+$^{197}$Au collisions even at low incident energies for all colliding geometries. We also see that $<A^{max}>$ at low energies is nearly independent of the impact parameter. This is due to the fact that at low incident energies, the available phase-space is too small to allow the frequent nucleon-nucleon collisions. But at high incident energies, the dynamics is dominated by the frequent nucleon-nucleon collisions that depend strongly on the impact parameter, therefore, stronger impact parameter dependence is observed for $<A^{max}>$ at higher incident energies. The emission of free nucleons exhibit just the opposite behaviour to that of $<A^{max}>$. We notice a gradual increase in the emission of free nucleons with incident energy for central and semi-central collisions. If we label the phenomenon with ≈ 60% matter as emitted nucleons as *vaporization* then we see the vaporization for central collisions above 150 MeV/nucleon. The rise and fall in the multiplicity of IMFs is seen for energy > 150 MeV/nucleon. The onset of multifragmentation is seen around 100 MeV/nucleon where IMF production is maximum.

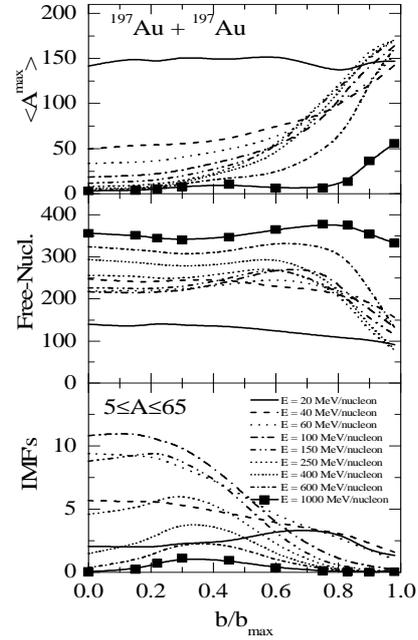

**Fig. 1** Different fragment multiplicities as a function of scaled impact parameter ($b/b_{max}$).

## Acknowledgments

This work is supported by a research grant from the Council of Scientific and Industrial Research (CSIR), Govt. of India, vide grant No. 09/135(0563)/2009-EMR-1.

## References:

[1] C. A. Ogilvie *et al.*, Phys. Rev. Lett. **67**, 1214 (1991).
[2] A. Schüttauf *et al.*, Nucl. Phys. A. **607**, 457 (1996).
[3] M. B. Tsang *et al.*, Phys. Rev. Lett. **71**, 1502 (1993).
[4] Y. K. Vermani and R. K. Puri, Eur. Phys. Lett. **85**, 62001 (2009); G. Peilert *et al.*, Phys. Rev. C **39**, 1402 (1989).
[5] R. K. Puri and S. Kumar, Phys. Rev. C **57**, 2744 (1998).
[6] J. Aichelin, Phys. Rep. **202,** 233 (1991).
[7] S. Goyal and R. K. Puri, Phys. Rev. C **83**, 047601 (2011).